\documentclass[prd,showpacs,amsmath,amssymb,superscriptaddress,nofootinbib,showkeys]{revtex4}
\usepackage{amssymb}
\usepackage{amsmath}
\usepackage[cp1251]{inputenc}
\usepackage[T2A]{fontenc}
\usepackage{graphicx}
\usepackage{floatflt}

\DeclareGraphicsExtensions{.pdf,.eps,.png,.jpg}

\begin{document}
\title{Magnetic Bianchi-I cosmology in the Horndeski theory}

\author{Ruslan K. Muharlyamov}
\email{rmukhar@mail.ru} \affiliation{Department of General
Relativity and Gravitation, Institute of Physics, Kazan Federal
University, Kremlevskaya str. 18, Kazan 420008, Russia}

\author{Tatiana N. Pankratyeva}
\email{ghjkl.15@list.ru} \affiliation{Department of Higher
Mathematics, Kazan State Power Engineering University,
Krasnoselskaya str. 51, Kazan 420066, Russia}

\author{Shehabaldeen O.A. Bashir}
\email{shehapbashir@gmail.com} \affiliation{Department of General
Relativity and Gravitation, Institute of Physics, Kazan Federal
University, Kremlevskaya str. 18, Kazan 420008, Russia; Department of physics, Faculty of Science, University of Khartoum, Khartoum, Sudan}


\begin{abstract}
We study the evolution of Bianchi-I space-times within the framework of the Horndeski theory with $G_5=\text{const}/X$. The space-times are filled a global unidirectional electromagnetic field interacting with a scalar  field. We consider the minimal interaction and the non-minimal interaction  by the law $f^2(\phi)F_{\mu\nu}F^{\mu\nu}$. The Horndeski theory allows anisotropy to grow over time, so the question arises of regulating the anisotropic level in this theory. Using  the designer method, we build models in which the anisotropic level tends to a small value as the Universe expands. One of the results is a model with a anisotropic bounce.

\end{abstract}



\maketitle

\section{Introduction}

Observations record magnetic fields at various scales of the Universe. Naturally, a large-scale magnetic field can influence cosmological evolution.
This influence has been the subject of study by many researchers over the years \cite{Doroshkevich,Thorne,Jacobs,Salimyx,Horwood,Bronnikov,Watanabe,Soda,Do,Nguyen}.  In this work, we adhere to the widespread hypothesis about the primordial origin of the magnetic field. According to the hypothesis, such a field could arise before or during primary inflation.
Therefore, it is interesting to study the interaction of the magnetic field with the scalar field, which causes inflation. The supposed connection of these fields provides great opportunities for observing the dark sector of the Universe.

The phenomenon of accelerated the Universe expansion  is the main motive for modifying the gravity theory. The Horndeski gravity (HG) \cite{Horndeski} is constructed in such a way that the motion equations are of the order of the derivative no higher than the second. In this sense, the HG is the most general variant of the scalar{tensor theory of gravitation. We use the following parametrization of the action density for the HG \cite{Kobayashi1}:
\begin{eqnarray}\label{lagr1} L_H=\sqrt{-g}\Big(\mathcal{L}_2+\mathcal{L}_3+\mathcal{L}_4+\mathcal{L}_5\Big) \,,\end{eqnarray}
$$\mathcal{L}_2 = G_2(\phi,X)\,,\, \mathcal{L}_3 = -
G_3(\phi,X)\Box\phi\,,$$
$$\mathcal{L}_4 = G_{4}(\phi,X) R +G_{4X}(\phi,X) \left[ (\square
\phi )^{2}-(\nabla_\mu \nabla_\nu \phi)^2  \right]\,,$$
\begin{equation} \mathcal{L}_5 = G_{5}(\phi,X) G_{\mu\nu}\,\nabla^\mu \nabla^\nu
\phi -\frac{1}{6}G_{5X}   \left[\left( \Box \phi \right)^3 -3 \Box
\phi (\nabla_\mu \nabla_\nu \phi)^2 + 2\left(\nabla_\mu \nabla_\nu
\phi \right)^3 \right], \label{lagr2}
\end{equation}
where $g$ is the determinant of metric tensor
$g_{\mu\nu}$; $R$ is the Ricci scalar and $G_{\mu\nu}$ is the
Einstein tensor; the factors $G_{i}$ ($i=2,3,4,5$) are arbitrary
functions of the scalar field $\phi$ and the canonical kinetic
term, $X=-\frac{1}{2}\nabla^\mu\phi \nabla_\mu\phi$. We consider
the definitions $G_{iX}\equiv \partial G_i/\partial X$,
 $(\nabla_\mu \nabla_\nu \phi)^2\equiv\nabla_\mu \nabla_\nu \phi
\,\nabla^\nu \nabla^\mu \phi$, and $\left(\nabla_\mu \nabla_\nu
\phi \right)^3\equiv \nabla_\mu \nabla_\nu \phi  \,\nabla^\nu
\nabla^\rho \phi \, \nabla_\rho \phi  \nabla^\mu \phi$.
We choose the electromagnetic part in the form
\begin{equation}\label{lagrF}\mathcal{L}_{F\phi}=-\frac{1}{4}[f(\phi)]^2F_{\mu\nu}F^{\mu\nu},\end{equation}
where $F_{\mu\nu}$ is the electromagnetic field. Option $f$=const corresponds to minimal interaction between the electromagnetic and the scalar fields.
 
 Modern high-quality instruments make it possible to obtain a wealth of observational cosmological data \cite{DESI1}. The observations of the Wilkinson Microwave Anisotropy Probe (WMAP) \cite{Hinshaw}, Planck satellites \cite{PAR} and the Dark Energy Spectroscopic Instrument (DESI) \cite{DESI2} say that the modern Universe is isotropic to a certain extent. However, anomalies are observed at large scale of the cosmic microwave background (CMB) radiation. This means that the early Universe could have been anisotropic. For example, in work \cite{Saadeh}  constraints were obtained on the isotropy of the Universe have been obtained in a general test using Planck’s data on the CMB temperature and polarization. The Bianchi Universe can explain these anomalies of the CMB \cite{Ellis, Komatsu, Thorsrud}.
The large-scale magnetic field vector specifies a  preferred spatial direction.  Therefore, it is natural to consider it in  an anisotropic cosmological model. Here we will limit ourselves to Bianchi type-I space-time (BI). BI  models have been studied from different perspectives \cite{Muharlyamov0,Amirhashchi0,Koussour,Koussour1,Akarsu,Sarmah,Hawking0,Hu,Momeni,Momeni1,Mehran,Petriakova}.
In the General relativity (GR), the scalar field is an "isotropic"\, object -- it does not generate anisotropy. In the HG, the "anisotropization"\, process \cite{SushkovStar, Tahara} by the scalar field is possible even in the absence of an anisotropic source. The "anisotropization"\, process is an increase of the anisotropic level  during the Universe expansion. But observational data say that the modern Universe is isotropic with a certain accuracy \cite{Rahman}. Therefore, the problem arises of regulating the development of anisotropy in these models.
To solve this problem we used  the designer method. Examples of this method can be found in \cite{Arjona,Muharlyamov1, Muharlyamov2, Muharlyamov3,Bernardo1,Bernardo2,Bernardo3}. Action densities (\ref{lagr1}) and (\ref{lagrF}) contain five arbitrary functions $G_i(X,\phi)$ and $f(\phi)$. Such a broad phenomenology expands the possibilities of  the  designer method.  The issue of isotropization in the modified theories of gravity is considered by many researchers \cite{Bhattacharya,Mandal,Alfedeel,Arora,Sharif,Esposito}.

In this article, for case $G_4=1/(16\pi)$, $G_5=\text{const}/X$, we develop a designer algorithm different from the previous ones  \cite{Muharlyamov1, Muharlyamov2, Muharlyamov3}.
We will apply this algorithm to the subclass of the HG:
\begin{equation} G_2=\varepsilon X-V(\phi),\,\, G_3=0. \end{equation}
The function $G_5(X,\phi)$ gives a non-minimal kinetic coupling to
the spacetime curvature \cite{Gao, Granda},
which may appear in some Kaluza-Klein theories \cite{Shafi1,
Shafi2}.
Using anisotropy constraints (relation $\sigma/H$), we recover potential $V(\phi)$ and function $f^2(\phi)$; $\sigma$ -- the shear scalar. Next, we study the features of the constructed models.

\section{Field equations}\label{sec2}

We  consider the homogeneous and anisotropic Bianchi I metric:
\begin{equation}
 ds^2 = -dt^2 + a^2_1(t)dx_1^2 + a^2_2(t)dx_2^2 + a^2_3(t)dx_3^2. \label{met0}
\end{equation}

As a result \cite{SushkovStar}, the corresponding field equations of the HG can be derived as

$$\frac{1}{8\pi}G^0_0= G_2 -
G_{2X}\dot{\phi}^2-3G_{3X}H\dot{\phi}^3+G_{3\phi}\dot{\phi}^2-
$$\begin{eqnarray} \label{00}
 - 5G_{5X}H_1H_2H_3\dot{\phi}^3 - G_{5XX}H_1H_2H_3\dot{\phi}^5+T^{({\rm em})0}_{0}\,,\end{eqnarray}
 $$\frac{1}{8\pi}G^{i}_{i} =G_2 -
\dot\phi \frac{d G_3}{dt} - $$\begin{eqnarray} \label{ii}-\frac{d}{dt}(G_{5X}\dot{\phi}^3
H_{j}H_{k})- G_{5X}\dot{\phi}^3 H_{j}H_{k}(H_{j}+H_{k})+T^{({\rm
em})i}_{i} \,,
\end{eqnarray}
$$\frac{1}{a^3}\frac{d}{dt}\left(a^3\dot{\phi}\, \Big[ G_{2X}-2G_{3\phi} +3H\dot\phi
G_{3X} +H_1H_2H_3(3G_{5X}\dot{\phi}
+G_{5XX}\dot{\phi}^3)\Big]\right)=$$\begin{eqnarray}\label{scalar0}=G_{2\phi} -\dot{\phi}^2(G_{3\phi\phi}
+G_{3X\phi}\ddot \phi)-\frac{f(\phi)f'_\phi
(\phi)}{2}F_{\gamma\delta}F^{\gamma\delta}\,,
\end{eqnarray}
\begin{eqnarray}\label{em} \partial_\mu[a^3 f^2(\phi)F^{\mu\nu}]=0\,.\end{eqnarray}
Here the dot denotes  the $t$-derivative, one has $H_i=\dot {
a}_i/{ a}_i$, and the average Hubble parameter is
$H=\dfrac{1}{3}\sum\limits_{i=1}^3 H_i\equiv \dot{ a}/{ a}$ with
${ a}=({ a}_1{ a}_2{ a}_3)^{1/3}$. In the equation (\ref{ii}) there is no summation over the indices $i$; the triples of indices $\{i,j,k\}$ take values $\{1,2,3\}$,
$\{2,3,1\}$, or $\{3,1,2\}$. The Einstein tensor components
are
\begin{eqnarray}
&&G^0_0=-\left(H_1H_2 +H_2H_3 +H_3H_1\right)\,, \\
&&G^{i}_{i}=-\left(\dot{H}_{j} +\dot{H}_{k} +H_{j}^2 +H_{k}^2
+H_{j}H_{k}\right)\,.
\end{eqnarray}

The stress--energy tensor of the electromagnetic field contains the interaction factor $f^2(\phi)$:
\begin{eqnarray}
T^{({\rm em})\mu}_{\nu}= f^2(\phi)\left(-\frac{1}{4}\delta^\mu_\nu
F_{\gamma\delta}F^{\gamma\delta} + F_{\nu \beta} F^{\mu
\beta}\right) \,. \label{em1}
\end{eqnarray}

We assume that there are homogeneous electric and magnetic fields
having the same direction  $x_3$.  Since the electromagnetic and scalar fields depend only on the time coordinate, system of equations (\ref{em}) takes the form $\partial_0[a^3 f^2(\phi)F^{03}]=0$. This equation has a solution $a^3 f^2(\phi)F^{03}=q_e$, where $q_e$ -- integration constant. The Bianchi identity
\begin{eqnarray}
\nabla_\mu F_{\nu\alpha}+\nabla_\alpha F_ {\mu\nu}+\nabla_\nu F_{\alpha\mu}=0
\end{eqnarray}
gives a solution $F_{21}=q_m$, where $q_m$ -- integration constant.
Thus the electromagnetic field tensor $F^{\gamma\delta}$ has non-vanishing components
\begin{eqnarray}
 F^{03}=-F^{30}=\frac{q_e}{a^3f^2(\phi)},\,\, F_{21}=-F_{12}=q_m\,. \label{em1}
\end{eqnarray}
 The electric and magnetic field strengths are determined by the equalities
\begin{eqnarray}\label{mag stren}
E^2=F_{03}F^{30}=\frac{q^2_e}{a^2_1a^2_2f^4(\phi)}\,,\, B^2=F_{21}F^{21}=\frac{q^2_m}{a^2_1a^2_2}\,.
\end{eqnarray}
The tensor $T^{({\rm em})\mu}_{\nu}$ has non-zero components:
$$T^{({\rm em})0}_{0}=T^{({\rm em})3}_{3}=-T^{({\rm em})1}_{1}=-T^{({\rm em})2}_{2}=$$\begin{eqnarray}=-\frac{f^2(\phi)}{2}(E^2+B^2)=
-\frac{\Psi(\phi)}{2a^2_1a^2_2}\,,\end{eqnarray}
where
\begin{eqnarray}\label{0Psi}
\Psi(\phi)\equiv\frac{q_e^2}{f^2(\phi)}+q_m^2f^2(\phi)>0\,.
\end{eqnarray}
We will assume that there is no the electric field:
\begin{eqnarray}\label{1Psi}
\Psi(\phi)=q_m^2f^2(\phi)>0\,.
\end{eqnarray}

Let's consider the following parametrization of three scalar factors:
\begin{equation}
ds^2 =
-dt^2+a^2(t)[e^{2(\beta_{+}+\sqrt{3}\beta_{-})}dx_1^2+e^{2(\beta_{+}-\sqrt{3}\beta_{-})}dx_2^2+e^{-4\beta_+}dx_3^2].
\label{met}
\end{equation}

The expansion rates in the direction $x_1$, $x_2$ and $x_3$ are given by
\begin{equation}
H_1=H+\dot{\beta}_{+}+\sqrt{3}\dot{\beta}_{-}\,,\,
H_2=H+\dot{\beta}_{+}-\sqrt{3}\dot{\beta}_{-}\,, \,
H_3=H-2\dot{\beta}_{+}\,. \label{H}
\end{equation}

From equations (\ref{00}), (\ref{ii}) and (\ref{scalar0}) we obtain the consequences
$$\frac{3}{8\pi}\big(H^2-\sigma^2\big)
=\frac{\Psi(\phi)}{2a^4e^{4\beta_+}}-G_2
+\dot{\phi}^2G_{2X}+3G_{3X}H\dot{\phi}^3
-G_{3\phi}\dot{\phi}^2+$$\begin{eqnarray}\label{1}+\dot{\phi}^3(5G_{5X}
+G_{5XX}\dot{\phi}^2) (H - 2\dot{\beta}_{+})
\big[(H+\dot{\beta}_+)^2 -3\dot{\beta}_{-}^2\big]\,,\end{eqnarray}

$$\frac{1}{8\pi}\big(2\dot{H}+3H^2 +
3\sigma^2\big)=-\frac{\Psi(\phi)}{6a^4e^{4\beta_+}}-G_2+G_{3\phi}\dot{\phi}^2
+G_{3X}\dot{\phi}^2\ddot{\phi}+$$\begin{eqnarray}\label{2}+\frac{d}{dt}\left[G_{5X}\dot{\phi}^3
\left(H^2-\sigma^2\right)\right]
+2G_{5X}\dot{\phi}^3\left(H^3+\dot{\beta}_{+}^3-3\dot{\beta}_{+}\dot{\beta}_{-}^2
\right)\,,\end{eqnarray}
\begin{eqnarray}\label{aniz1}\frac{\dot{\beta}_{-}}{8\pi} +G_{5X}\dot\phi^3
\left(2\dot\beta_{+}\dot\beta_{-} -H\dot\beta_{-}\right) =
\frac{C_-}{{ a}^3}\,,\end{eqnarray}
\begin{eqnarray}\label{aniz2}\frac{\dot{\beta}_{+}}{8\pi} +G_{5X}\dot\phi^3\left(\dot\beta_{-}^2-\dot\beta_{+}^2
-H\dot\beta_{+}\right)=\frac{1}{3a^3}\int\frac{\Psi(\phi)dt}{ae^{4\beta_+}}+\frac{C_+}{a^3},\end{eqnarray}
$$\dot{\phi}\, \Big\{ G_{2X}-2G_{3\phi}+3H\dot\phi
G_{3X}
+(3G_{5X}\dot{\phi}
+G_{5XX}\dot{\phi}^3)(H - 2\dot{\beta}_{+})\times$$\begin{eqnarray}\label{scalar}
\times\big[(H+\dot{\beta}_+)^2 -3\dot{\beta}_{-}^2\big]\Big\}=
\frac{1}{a^3}\int \left(G_{2\phi}-\dot{\phi}^2(G_{3\phi\phi}
+G_{3X\phi}\ddot \phi)-\frac{\Psi'_\phi}{2a^4e^{4\beta_+}}\right)a^3dt+\frac{C_\phi}{a^3}\,,\end{eqnarray}
\begin{eqnarray}\label{1234}\sigma^2 \equiv\dot{\beta}^2_{+} + \dot{\beta}^2_{-},\end{eqnarray}
where $C_\phi$, $C_+$ and $C_-$ are integration constants.

The theory with $G_{5X}\neq0$ gives the system of nonlinear equations
(\ref{aniz1}), (\ref{aniz2}) for $\dot\beta_{\pm}$.  Due to the nonlinearity of  equations (\ref{aniz1}), (\ref{aniz2}), the "anisotropization"\, process \cite{SushkovStar, Tahara} by the scalar field is possible even in the absence of an anisotropic source $\dfrac{\Psi(\phi)}{ae^{4\beta_+}}$.

Further, we put
\begin{equation}C_-=C_+=0\,.\end{equation}
Then, resolving (\ref{aniz1}) and (\ref{aniz2}) with respect to $\dot{\beta}_{\pm}$, we obtain one of the solutions
\begin{equation}\label{b1}\dot\beta_{+}=\frac{1}{2}\left(H-\frac{1}{8\pi \cdot
G_{5X}\dot{\phi}^3}\right),\end{equation}
\begin{equation}\label{b0}\dot\beta_{-}=\sqrt{\frac{3}{4}\left(H-\frac{1}{8\pi \cdot
G_{5X}\dot{\phi}^3}\right)^2+\frac{a^{-3}}{3G_{5X}\dot{\phi}^3}\int dt\, \frac{\Psi(\phi)}{ae^{4\beta_+}}}\,\,.\end{equation}

In view of (\ref{b1}) and (\ref{b0}), from equations
(\ref{1}) and (\ref{scalar}) we obtain
$$3H\left\{G_{3X}\dot{\phi}^3+\frac{1}{(8\pi)^2 \cdot G_{5X}\dot{\phi}^3}\left[3+\frac{G_{5XX}\cdot\dot{\phi}^2}{G_{5X}}\right]\right\}
=$$$$=-\frac{\Psi(\phi)}{2a^4e^{4\beta_+}}+\frac{a^{-3}}{8\pi G_{5X}\dot{\phi}^3}\left[4+\frac{G_{5XX}\cdot\dot{\phi}^2}{G_{5X}}\right]\int dt\, \frac{\Psi}{ae^{4\beta_+}}+$$
\begin{equation}\label{G5grav}+G_{3\phi}\dot{\phi}^2+G_2-\dot{\phi}^2G_{2X}+\frac{1}{(8\pi)^3
\cdot(
G_{5X}\dot{\phi}^3)^2}\left[7+\frac{2G_{5XX}\cdot\dot{\phi}^2}{G_{5X}}\right],\end{equation}
$$3H\left\{G_{3X}\dot{\phi}^3+\frac{1}{(8\pi)^2 \cdot G_{5X}\dot{\phi}^3}\left[3+\frac{G_{5XX}\cdot\dot{\phi}^2}{G_{5X}}\right]\right\}
=-\dot{\phi}^2G_{2X}+2G_{3\phi}\dot{\phi}^2+\frac{C_\phi \dot{\phi}}{a^3}+$$$$+\frac{2}{(8\pi)^3 \cdot(
G_{5X}\dot{\phi}^3)^2}\left[3+\frac{G_{5XX}\cdot\dot{\phi}^2}{G_{5X}}\right]+\frac{a^{-3}}{8\pi G_{5X}\dot{\phi}^3}\left[3+\frac{G_{5XX}\cdot\dot{\phi}^2}{G_{5X}}\right]\times$$\begin{equation}\label{G5scalar}
\times\int dt\, \frac{\Psi}{ae^{4\beta_+}}+
\frac{\dot{\phi}}{a^3}\int dt\, a^3\left(G_{2\phi}-\dot{\phi}^2(G_{3\phi\phi}
+G_{3X\phi}\ddot \phi)-\frac{\Psi'_\phi}{2a^4e^{4\beta_+}}\right).\end{equation}
The  equation (\ref{2}) can be ignored, since it is automatically
fulfilled by virtue of the Bianchi identities.

Next, let's put
\begin{equation}\label{G5}G_5=-\frac{1}{32\pi \gamma X}\,.\end{equation}
The choice (\ref{G5}) allows, without unnecessary technical difficulties, to obtain interesting cosmological models based on the nonlinearity of equations (\ref{aniz1}), (\ref{aniz2}) relative to $\dot\beta_{\pm}$. As will be shown below, the function (\ref{G5}) allows for acceptable anisotropy behavior in cosmological models. From (\ref{b1}) and (\ref{b0}) it follows
\begin{equation}\label{bb}\dot\beta_{+}=\frac{1}{2}\left(H-\gamma\dot{\phi}\right)\Rightarrow \beta_+=
\frac{1}{2}\left(\ln a-\gamma\phi\right)+\text{const}\,,\end{equation}
\begin{equation}\label{a1a2qur}\frac{1}{a_1^2a_2^2}=\frac{1}{a^4e^{4\beta_+}}=\frac{c_0e^{2\gamma \phi}}{a^6},\,\, c_0>0\,,\end{equation}
\begin{equation}\label{bb1}\dot\beta_{-}=\sqrt{\frac{3}{4}\left(H-\gamma\dot{\phi}\right)^2+\frac{8\pi\gamma\dot{\phi}}{3a^{3}}
\int dt\, \frac{c_0e^{2\gamma\phi}\Psi(\phi)}{a^3}}\,\,.\end{equation}

The shear scalar $\sigma^2$ and the Hubble parameters take the form
\begin{equation}\label{sigma}\sigma^2=\left(H-\gamma\dot{\phi}\right)^2+\frac{8\pi\gamma\dot{\phi}}{3a^{3}}
\int dt\, \frac{c_0e^{2\gamma\phi}\Psi(\phi)}{a^3}.\end{equation}
$$H_{1,2}=\frac{1}{2}\left(3H-\gamma\dot{\phi}\right)\pm$$
\begin{equation}\label{h12aa}\pm\sqrt{\frac{9}{4}\left(H-\gamma\dot{\phi}\right)^2+\frac{8\pi\gamma\dot{\phi}}{a^{3}}
\int dt\, \frac{c_0e^{2\gamma\phi}\Psi(\phi)}{a^3}}\,\,,\end{equation}

\begin{equation}\label{h3aab}H_3=\gamma\dot{\phi}\,.\end{equation}

The system (\ref{G5grav})-(\ref{G5scalar}) becomes simple:
\begin{equation}\label{G5grav1}3H\left\{G_{3X}\dot{\phi}^3-\frac{\gamma\dot{\phi}}{8\pi}\right\}
=G_2-\dot{\phi}^2G_{2X}+G_{3\phi}\dot{\phi}^2-\frac{\gamma^2\dot{\phi}^2}{8\pi}-\frac{c_0e^{2\gamma\phi}}{2a^6}\Psi(\phi)
\,,\end{equation}
$$3H\left\{G_{3X}\dot{\phi}^3-\frac{\gamma\dot{\phi}}{8\pi}\right\}
=-\dot{\phi}^2G_{2X}+2G_{3\phi}\dot{\phi}^2-\frac{2\gamma^2\dot{\phi}^2}{8\pi}+\frac{C_\phi \dot{\phi}}{a^3}+$$
\begin{equation}\label{G5scalar1}+
\frac{\dot{\phi}}{a^3}\int dt\, a^3\left[G_{2\phi}-\dot{\phi}^2(G_{3\phi\phi}
+G_{3X\phi}\ddot \phi)-
\frac{c_0e^{2\gamma\phi}}{2a^6}\left(\Psi'_\phi+2\gamma\Psi\right)\right]\,.\end{equation}

Next, we use  the designer method.
We have five unknown functions $\{a(t), \phi(t), G_2(X,\phi),
G_{3}(X,\phi), \Psi(\phi)\}$ and two independent equations. Therefore, we have three degrees of freedom.  They can be applied in different ways. One way is to choose the type of function's dependence on its arguments. For example, $G_2(X,\phi)=\pm X-Ae^{\lambda\phi}$.
Another option is to impose coupling on somewhat undefined functions. Here we will assume such the coupling
\begin{equation}\label{1int}-G_{2\phi}+\dot{\phi}^2(G_{3\phi\phi}
+G_{3X\phi}\ddot \phi)+\frac{c_0e^{2\gamma\phi}}{2a^6}\left(\Psi_\phi+2\gamma\Psi\right)=H\frac{U'_a(a)}{a^2}\,.\end{equation}
This combination is included in the integral  in (\ref{G5scalar1}). The integral is transformed as follows
$$\int dt\, a^3\left[G_{2\phi}-\dot{\phi}^2(G_{3\phi\phi}
+G_{3X\phi}\ddot \phi)-
\frac{c_0e^{2\gamma\phi}}{2a^6}\left(\Psi'_\phi+2\gamma\Psi\right)\right]=$$\begin{equation}\label{G5scalar12}=-\int dt\, a^3\cdot H\frac{U'_a(a)}{a^2}=-U(a).\end{equation}
In view of (\ref{G5scalar12}), the equation (\ref{G5scalar1}) will be rewritten:
\begin{equation}\label{G5scalar11}3H\left\{G_{3X}\dot{\phi}^3-\frac{\gamma\dot{\phi}}{8\pi}\right\}
=-\dot{\phi}^2G_{2X}+2G_{3\phi}\dot{\phi}^2-\frac{2\gamma^2\dot{\phi}^2}{8\pi}+\frac{ \dot{\phi}}{a^3}\left(C_\phi-U(a)\right)
\,.\end{equation}
By specifying function $U(a)$, we will finally define the coupling (\ref{1int}) and thereby we exhaust one degree of freedom. We can look at equation (\ref{1int}) as a way to implicitly regulate the non-minimal interaction $\Psi(\phi)$ of the scalar field and the electromagnetic field through function $U(a)$.
Thus, we get system (\ref{G5grav1}), (\ref{1int}), (\ref{G5scalar11}).

\section{Model $G_{3}=0$, $G_2=\varepsilon X-V$}

Let's consider the following model
\begin{equation}\label{G3G2}G_{3}=U(a)=C_\phi=0,\,\, G_2=\varepsilon X-V,\,\,\varepsilon=\pm1\,.\end{equation}
In this case, from equation (\ref{G5scalar11}) it follows
\begin{equation}\label{Hgamma}\frac{3H\gamma}{8\pi}=\dot{\phi}\left(\varepsilon+\frac{2\gamma^2}{8\pi}\right).\end{equation}
Therefore, from (\ref{h3aab}) we get
\begin{equation}\label{h3a}\frac{H_3}{H}=\frac{3\gamma^2}{8\pi\left(\varepsilon+\frac{2\gamma^2}{8\pi}\right)}=\text{const},\,\,
\frac{\dot\beta_{+}}{H}=\frac{1}{2}\cdot\frac{\varepsilon-\frac{\gamma^2}{8\pi}}{\varepsilon+\frac{2\gamma^2}{8\pi}}=\text{const}.\end{equation}

An important criterion for the viability of any anisotropic model is a sufficiently low level of anisotropy at certain stages of the Universe evolution. In particular, it was argued in \cite{Hawking} that, from the point of view of the particle production, a significant decrease in anisotropy should occur quite early, no later than the beginning of primary nucleosynthesis.  The work  \cite{Campanelli} analyzed the effects caused by cosmic anisotropy on the primordial production of $^4$He.

The parameter $\sigma^2/H^2=(\dot{\beta}^2_{+} + \dot{\beta}^2_{-})/H^2$ describes the anisotropy and the isotropization process of the Universe. For example, when studying anisotropic inflation, from the Planck data they give the current estimate $\sigma/H\sim 10^{-9}$ \cite{Kim}. Constant ratios (\ref{h3a}) mean that at any moment of time there will be a constant component in the anisotropy ($\sigma^2/H^2$). We put
\begin{equation}\label{malaniz}\varepsilon=1,\,\, \frac{8\pi}{\gamma^2}=1+\mu,\,\, |\mu|=\text{const}\ll 1,\end{equation}
 then this component will be small:
\begin{equation}\label{malaniz01}\left|\frac{\dot\beta_{+}}{H}\right|=\frac{1}{2}\cdot\left|\frac{\mu}{3+\mu}\right|\ll 1
\,\, \, \text{or} \,\,\, \frac{H_3}{H}=\frac{3}{3+\mu}\approx 1.\end{equation}
In the model under consideration, estimate (\ref{malaniz01}) is valid on all time scales. Limitation (\ref{malaniz})   is a necessary condition, but not sufficient for the smallness of $\sigma/H$ on certain time scales, since there is also a component $\dot\beta_{-}^2/H^2$. This ratio may depend on time.

Functions $\beta_{+}$, $\dot\beta_{-}$, $\sigma^2/H^2$ and $H_{i}$ take the form
\begin{equation}\label{betplus}\beta_{+}=\frac{1}{2}\cdot\frac{\frac{8\pi}{\gamma^2}-1}{\frac{8\pi}{\gamma^2}+2}\cdot\ln a+\text{const}\,,\end{equation}
\begin{equation}\label{betminus}\dot\beta_{-}=\sqrt{\frac{3}{4}H^2\cdot
\left(\frac{\frac{8\pi}{\gamma^2}-1}{\frac{8\pi}{\gamma^2}+2}\right)^2+
\frac{8\pi H}{\left(2+\frac{8\pi}{\gamma^2}\right)a^{3}}
\cdot\int dt\, \frac{c_0e^{2\gamma\phi}\Psi(\phi)}{a^3}}\,,\end{equation}
\begin{equation}\label{sigmanaH}\frac{\sigma^2}{H^2}=
\left(\frac{\frac{8\pi}{\gamma^2}-1}{\frac{8\pi}{\gamma^2}+2}\right)^2+
\frac{8\pi}{\left(2+\frac{8\pi}{\gamma^2}\right)Ha^{3}}
\cdot\int dt\, \frac{c_0e^{2\gamma\phi}\Psi(\phi)}{a^3}\,,\end{equation}

\begin{equation}\label{ha1}H_3=H\cdot \frac{3}{2+\frac{8\pi}{\gamma^2}}\,,\end{equation}

$$H_{1,2}=\frac{3}{2}H\cdot\frac{1+\frac{8\pi}{\gamma^2}}{2+\frac{8\pi}{\gamma^2}}\pm$$
\begin{equation}\label{h12a}\pm\sqrt{\frac{9}{4}H^2\cdot
\left(\frac{\frac{8\pi}{\gamma^2}-1}{\frac{8\pi}{\gamma^2}+2}\right)^2+
\frac{24\pi H}{\left(2+\frac{8\pi}{\gamma^2}\right)a^{3}}
\cdot\int dt\, \frac{c_0e^{2\gamma\phi}\Psi(\phi)}{a^3}}\,.\end{equation}
  As you can see in (\ref{sigmanaH}), there is a constant small term $\left(\frac{\frac{8\pi}{\gamma^2}-1}{\frac{8\pi}{\gamma^2}+2}\right)^2=\frac{\mu^2}{(3+\mu)^2}\ll 1$. This model is anisotropic at any time. It is necessary to investigate the behavior of term
\begin{equation}\frac{8\pi}{\left(2+\frac{8\pi}{\gamma^2}\right)Ha^{3}}
\cdot\int dt\, \frac{c_0e^{2\gamma\phi}\Psi(\phi)}{a^3}\end{equation}
for a complete analysis of anisotropy.

System (\ref{G5grav1}), (\ref{1int}), (\ref{G5scalar11}) has consequences:
\begin{equation}\label{phs1}\phi=\ln\left(\frac{a}{c_1}\right)^{\frac{3}{\gamma\left(2+\frac{8\pi}{\gamma^2}\right)}}\,,\end{equation}
\begin{equation}\label{VPsi}  V'_{\phi}+\frac{c_0}{2c_1^6}\exp\left[-2\gamma\phi\left(1+\frac{8\pi}{\gamma^2}\right)\right]
\left(\Psi'_\phi+2\gamma\Psi\right)=0\,,\end{equation}
\begin{equation}\label{dotphi} H^2=\frac{16\pi}{9}\left(2+\frac{8\pi }{\gamma^2}\right)\left[V+
\frac{c_0}{2c_1^6}\Psi\exp\left\{-2\gamma\phi\left(1+\frac{8\pi }{\gamma^2}\right)\right\}\right]\,.\end{equation}
Of the three, one degree of freedom remains.

From equalities  (\ref{mag stren}),  (\ref{a1a2qur}) and (\ref{phs1}), it follows that the magnetic field strength decreases monotonically with the Universe expansion:
\begin{equation}B=\text{const}\cdot a^{-3\cdot\frac{1+\frac{8\pi }{\gamma^2}}{2+\frac{8\pi }{\gamma^2}}}\approx \text{const}\cdot a^{-2}\,.\end{equation}

\subsection{Model without magnetic field}

In the absence of the magnetic field, $\Psi=0$, from (\ref{VPsi}) we obtain a constant potential $V=V_0$.
All Hubble parameters are constant (see (\ref{ha1}), (\ref{h12a}) and (\ref{dotphi})). The Universe is expanding with acceleration in all directions:
$$H=H_0\equiv\frac{4}{3}\sqrt{\pi\left(2+\frac{8\pi }{\gamma^2}\right) V_0}>0\Rightarrow a(t)=
a_0e^{H_0t},$$
$$H_1=3H_0\cdot\frac{8\pi}{\gamma^2\left(2+\frac{8\pi}{\gamma^2}\right)}>0\Rightarrow a_1(t)=\text{const}\cdot a^{3\cdot\frac{8\pi}{\gamma^2\left(2+\frac{8\pi}{\gamma^2}\right)}},$$
\begin{equation}\label{bezMh123}H_2=H_3=3H_0\cdot\frac{1}{2+\frac{8\pi}{\gamma^2}}>0\Rightarrow a_2(t)=a_3(t)=\text{const}\cdot a^{\frac{3}{2+\frac{8\pi}{\gamma^2}}}.\end{equation}
Two scale factors are equal ($a_2(t)=a_3(t)$), that is, this is the locally rotationally symmetric (LRS) BI model.
From (\ref{sigmanaH}) it follows that the model has a constant level of anisotropy:
\begin{equation}\label{bezsi}\frac{\sigma^2}{H^2}=\left(\frac{\frac{8\pi}{\gamma^2}-1}{\frac{8\pi}{\gamma^2}+2}\right)^2=\text{const}.\end{equation}
As a consequence of assumption (\ref{malaniz}), the level of anisotropy is low:
\begin{equation}\frac{\sigma^2}{H^2}\approx\frac{\mu^2}{9}\ll1,\, H_2=H_3\approx H_1\approx H_0.\end{equation}

Scalar field $\phi$ linear with respect to cosmological time:
\begin{equation}\phi=\frac{3H_0 t}{\gamma\left(2+\frac{8\pi}{\gamma^2}\right)}+\text{const}\,.\end{equation}

\subsection{Minimal coupling with the magnetic field}
In the case of minimal interaction with the magnetic field:
\begin{equation}\Psi=q^2,\end{equation}
we get the potential from (\ref{VPsi}):
\begin{equation}V=V_0+\frac{c_0q^2}{2c_1^6\left(1+\frac{8\pi }{\gamma^2}\right)}\exp\left[-2\gamma\phi\left(1+\frac{8\pi }{\gamma^2}\right)\right]\,.\end{equation}
Then
\begin{equation}\label{minH2}H^2=\frac{16\pi}{9}\left(2+\frac{8\pi }{\gamma^2}\right)
\left[V_0+\frac{c_0q^2\left(2+\frac{8\pi}{\gamma^2}\right)}{2c_1^6\left(1+\frac{8\pi}{\gamma^2}\right)}
\left(\frac{c_1}{a}\right)^{6\cdot\frac{1+\frac{8\pi }{\gamma^2}}{2+\frac{8\pi }{\gamma^2}}}\right]\,.\end{equation}
The scale factor $a(t)$, $H(t)$ and the deceleration
parameter (DP) are, respectively, given by
\begin{equation}\label{scal}a=c^{-\frac{1}{1+\frac{8\pi }{\gamma^2}}}_1\left\{\frac{c_0q^2\left(2+\frac{8\pi}{\gamma^2}\right) }{2V_0\left(1+\frac{8\pi}{\gamma^2}\right)}\sinh^2\left[ t/t_*\right]
\right\}^{\frac{2+\frac{8\pi}{\gamma^2}}{6\left(1+\frac{8\pi}{\gamma^2}\right)}},\,\, t\geq 0\,\Rightarrow\end{equation}
\begin{equation}\label{minH21}\Rightarrow H=\frac{2+\frac{8\pi}{\gamma^2}}{3t_*\left(1+\frac{8\pi}{\gamma^2}\right)}\coth\left[ t/t_*\right],\,\, \frac{1}{t^*}\equiv
4\left(1+\frac{8\pi }{\gamma^2}\right)\sqrt{\frac{\pi V_0}{2+\frac{8\pi}{\gamma^2}}}\,,\end{equation}
\begin{equation}\label{DP}q_d=\frac{d}{dt}\frac{1}{H}-1=3\cdot\frac{1+\frac{8\pi}{\gamma^2}}{2+\frac{8\pi}{\gamma^2}}\cdot\cosh^{-2}\left( t/t_*\right)-1\,.\end{equation}
Scalar field $\phi$ has the following time dependence
\begin{equation}\phi=\frac{1}{\gamma\left(1+\frac{8\pi }{\gamma^2}\right)}\ln \left[\sqrt{\frac{c_0q^2\left(2+\frac{8\pi}{\gamma^2}\right)}{2V_0c_1\left(1+\frac{8\pi}{\gamma^2}\right)}}\sinh( t/t_*)\right]\,.\end{equation}

\begin{figure}[h]
\includegraphics[width=10cm]{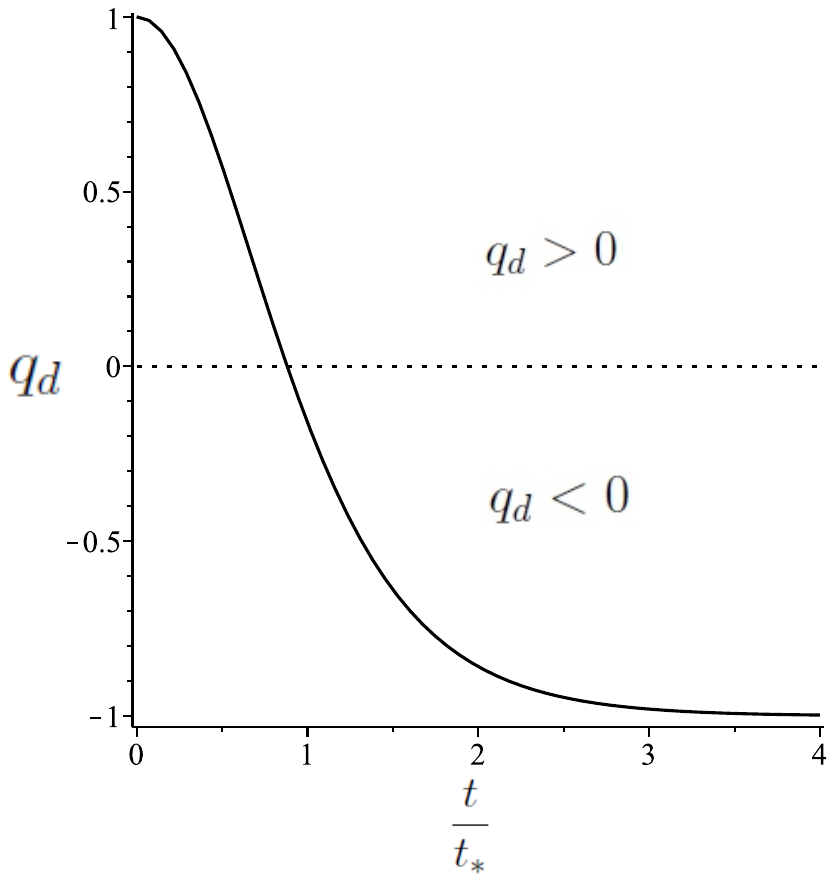}
\caption{The profile of the DP, $\mu=10^{-4}$. \label{risDP}}
\end{figure}

Let's consider the "geometric mean"\, behavior of the Universe. Based on the type of scale factor (\ref{scal}), the Universe is expanding all the time. From Fig. \ref{risDP} it is clear that the DP (\ref{DP}) evolves from positive values at past epoch to negative values at late time. In this model, there are two phases of the evolution of the Universe.
In the first phase, there is no acceleration ($q_d\geq0$), and the second one is characterized by the acceleration expansion of the Universe ($q_d<0$). For $t/t_*\ll 1$ we have the approximation $H^2\sim a^{-6\cdot\frac{1+\frac{8\pi }{\gamma^2}}{2+\frac{8\pi }{\gamma^2}}}$, or $H\sim\frac{2+\frac{8\pi }{\gamma^2}}{3\left(1+\frac{8\pi}{\gamma^2}\right)}\cdot 1/t$. In view of (\ref{malaniz}), the degree is approximately equal
\begin{equation}-6\cdot\frac{1+\frac{8\pi }{\gamma^2}}{2+\frac{8\pi }{\gamma^2}}\approx
-4\left(1+\frac{\mu}{6}\right)\approx -4\,.\end{equation}
This corresponds to radiation with the equation of state $w\approx 1/3$.  In late-times $(t/t_*\gg1)$, the DP tends to $-1$, which corresponds to de Sitter type expansion:
\begin{equation}H\rightarrow H_0=\frac{4}{3}\sqrt{\pi\left(2+\frac{8\pi }{\gamma^2}\right) V_0}\,.\end{equation}

Next, we will consider the anisotropic properties of the model. From (\ref{sigmanaH}) and (\ref{h12a}) it follows
$$\frac{\sigma^2}{H^2}=
\left(\frac{\frac{8\pi}{\gamma^2}-1}{\frac{8\pi}{\gamma^2}+2}\right)^2+\frac{3}{\left(2+\frac{8\pi}{\gamma^2}\right)^2}
\times$$\begin{equation}\label{sigmanaHmin}\times\frac{\sinh^{-\frac{1}{1+\frac{8\pi}{\gamma^2}}}\left[ t/t_*\right]
}{\cosh\left[ t/t_*\right]}\int d\left[ t/t_*\right] \cdot \sinh^{-\frac{\frac{8\pi}{\gamma^2}}{1+\frac{8\pi}{\gamma^2}}}\left[ t/t_*\right]
\,,\end{equation}
$$H_{1,2}=\frac{3}{2}H\cdot\frac{1+\frac{8\pi}{\gamma^2}}{2+\frac{8\pi}{\gamma^2}}\pm \frac{3}{2}H\times$$
\begin{equation}\label{minh12a}\times\left\{
\left(\frac{\frac{8\pi}{\gamma^2}-1}{\frac{8\pi}{\gamma^2}+2}\right)^2+
\frac{4\sinh^{-\frac{1}{1+\frac{8\pi}{\gamma^2}}}\left[ t/t_*\right]
}{\left(2+\frac{8\pi}{\gamma^2}\right)^2\cosh\left[ t/t_*\right]}\int d\left[ t/t_*\right] \cdot \sinh^{-\frac{\frac{8\pi}{\gamma^2}}{1+\frac{8\pi}{\gamma^2}}}\left[ t/t_*\right]\right\}^{1/2}\,.\end{equation}
Parameter $H_3$ remains the same (\ref{ha1}).

We mark the necessary properties of the function:
\begin{equation}\frac{\sinh^{-\frac{1}{1+\frac{8\pi}{\gamma^2}}}\left[ t/t_*\right]
}{\cosh\left[ t/t_*\right]}\int d\left[ t/t_*\right] \cdot \sinh^{-\frac{\frac{8\pi}{\gamma^2}}{1+\frac{8\pi}{\gamma^2}}}\left[ t/t_*\right]\rightarrow 1+\frac{8\pi}{\gamma^2},\,\, \text{as} \,\, \,t/t_*\rightarrow 0,\end{equation}
\begin{equation}\label{rem1}\frac{\sinh^{-\frac{1}{1+\frac{8\pi}{\gamma^2}}}\left[ t/t_*\right]
}{\cosh\left[ t/t_*\right]}\int d\left[ t/t_*\right] \cdot \sinh^{-\frac{\frac{8\pi}{\gamma^2}}{1+\frac{8\pi}{\gamma^2}}}\left[ t/t_*\right]\propto \exp(-2t/t_*),\,\, \text{as}\,\, \,t/t_*\rightarrow+\infty.\end{equation}

\begin{figure}[h]
\includegraphics[width=10cm]{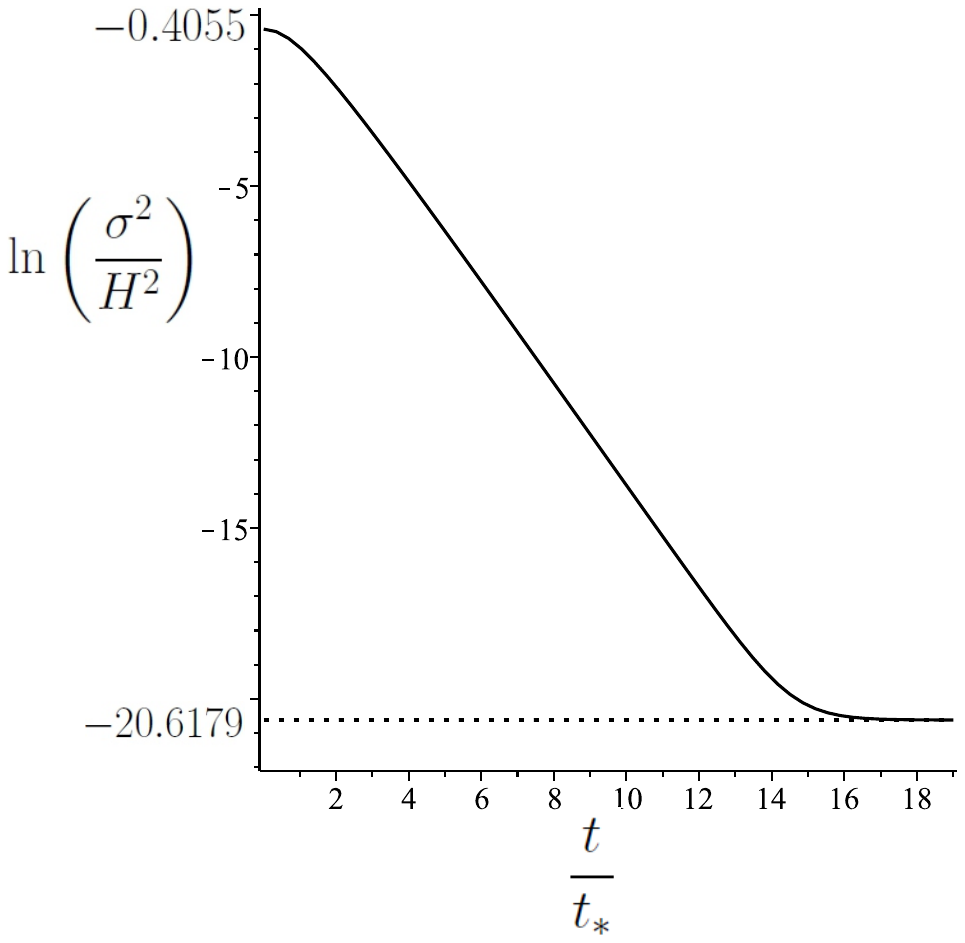}
\caption{The profile of the $\ln(\sigma^2/H^2)$, $\mu=10^{-4}$. \label{risS_H}}
\end{figure}

As seen in Fig. \ref{risS_H}, the parameter $\sigma^2/H^2$ tends monotonically to a constant value over time. It is the bounded function:
\begin{equation}\left(\frac{\frac{8\pi}{\gamma^2}-1}{\frac{8\pi}{\gamma^2}+2}\right)^2\leq\frac{\sigma^2}{H^2}\leq
\frac{1}{\left(2+\frac{8\pi}{\gamma^2}\right)^2}
\left[\left(\frac{8\pi}{\gamma^2}-1\right)^2+
3\left(1+\frac{8\pi}{\gamma^2}\right)\right]\,,\,\, \text{or}\end{equation}
\begin{equation}\frac{\mu^2}{9}\lesssim\frac{\sigma^2}{H^2}\lesssim\frac{2}{3}\left(1-\frac{\mu}{6}\right)\,.\end{equation}
The anisotropy of the Universe is decreasing, $\sigma^2/H^2\rightarrow \mu^2/9\ll 1$, $t/t_*\rightarrow +\infty$.  According to (\ref{sigmanaHmin}) and (\ref{rem1}), the decrease in anisotropy occurs quickly, according to the exponential law.

\begin{figure}[h]
\includegraphics[width=10cm]{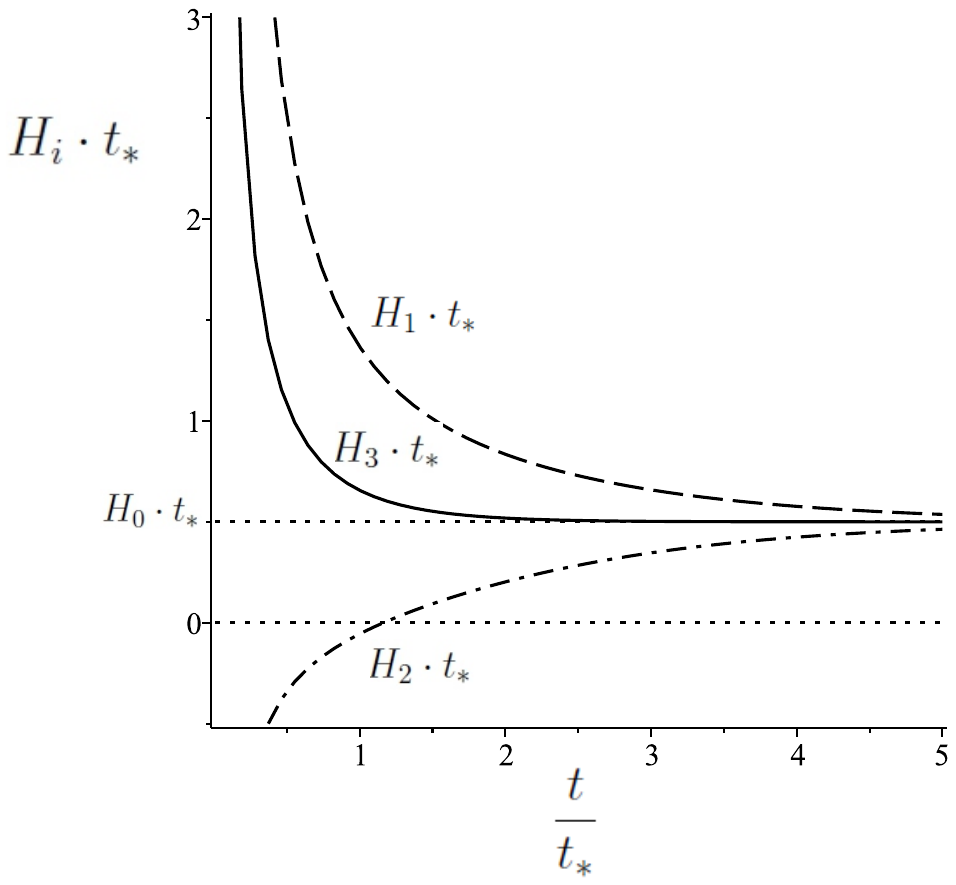}
\caption{The profile of the $H_i\cdot t_*$; $H_0\cdot t_*=\frac{2+\frac{8\pi}{\gamma^2}}{3\left(1+\frac{8\pi}{\gamma^2}\right)}$, $\mu=10^{-4}$. \label{risH123}}
\end{figure}

As seen in Fig. \ref{risH123}, the functions $H_i$ tend monotonically to  constant values over time, $t/t_*\gg 1$:
\begin{equation}H_1\rightarrow3H_0\cdot\frac{8\pi}{\gamma^2\left(2+\frac{8\pi}{\gamma^2}\right)};\,\,
H_2, H_3\rightarrow3H_0\cdot\frac{1}{2+\frac{8\pi}{\gamma^2}}\,,\end{equation}
that is, in this approximation we obtain model (\ref{bezMh123}).

The Universe is expanding along the $x_1$ and $x_3$ axes: $H_{1,3}>0$.
The Hubble parameter $H_2$ takes on negative and positive values. The Universe contracts in early times, and expands in late times along axis $x_2$.
At early time, $t/t_*\ll 1$:
\begin{equation}H_{1,2}\propto \frac{1}{2t}\left\{1\pm\left[1+\frac{4}{\left(2+\frac{8\pi}{\gamma^2}\right)^2}\right]^{1/2}\right\}\approx\frac{1\pm\sqrt{2}}{2t},\,\, H_2<0\,,\end{equation}
\begin{equation}H_3\propto\frac{1}{t\left(1+\frac{8\pi}{\gamma^2}\right)}\approx\frac{1}{2t}\,.\end{equation}
At early times  scalar factors have the approximation:
\begin{equation}a_{1,2}\propto \left(\frac{t}{t_*}\right)^{\frac{1}{2}\left\{1\pm\left[1+\frac{4}{\left(2+\frac{8\pi}{\gamma^2}\right)^2}\right]^{1/2}\right\}},\,\,
a_3\propto\left(\frac{t}{t_*}\right)^{\frac{1}{1+\frac{8\pi}{\gamma^2}}}\,,\,\, \text{or}\end{equation}
\begin{equation}a_{1}(a)\sim a^{p_1},\,\, a_2(a)\sim\frac{1}{a^{|p_2|}},\,\, a_3(a)=\text{const}\cdot a^{\frac{3}{2+\frac{8\pi}{\gamma^2}}}\,,\end{equation}
where
\begin{equation}p_{1,2}\equiv\frac{3}{2\left(2+\frac{8\pi}{\gamma^2}\right)}\left[1+\frac{8\pi}{\gamma^2}\pm
\left\{\left(\frac{8\pi}{\gamma^2}+1\right)^2+4\right\}^{1/2}\right]\approx 1\pm\sqrt{2}\,.\end{equation}

\begin{figure}[h]
\includegraphics[width=10cm]{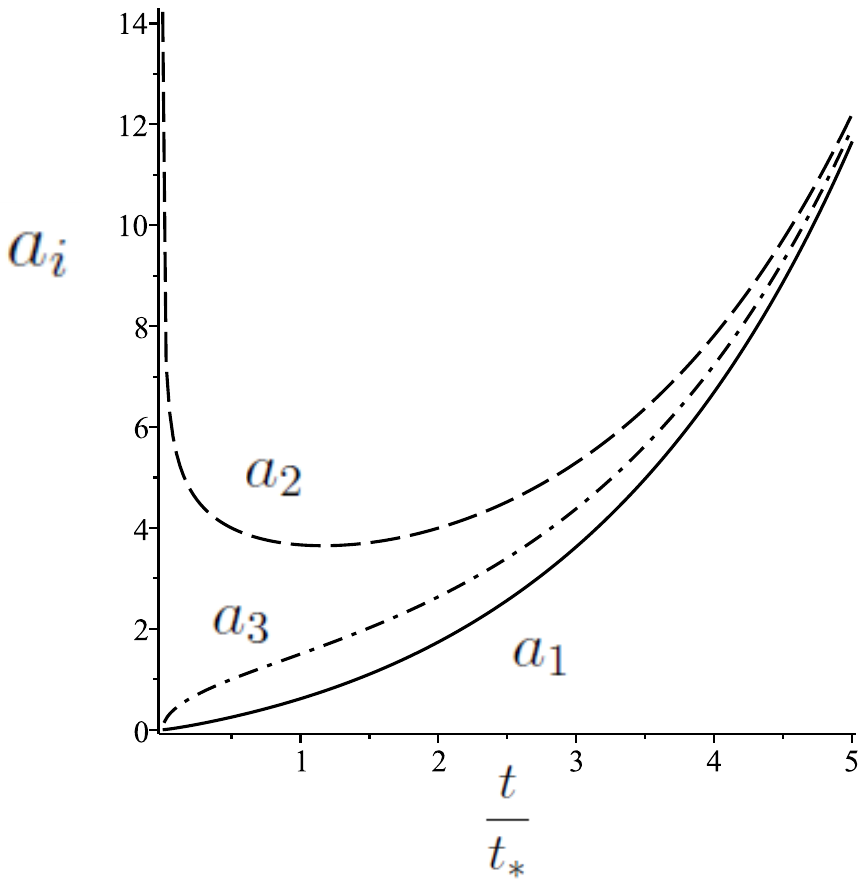}
\caption{The profile of the $a_i(t/t_*)$ with initial conditions $a_1(0.5)=1/4$, $a_2(0.5)=4$, $a_3(0.5)=1$, $\mu=10^{-4}$. \label{risa123}}
\end{figure}

Scale factors $a_i(t)$ are shown in Fig. \ref{risa123}. The figure shows a bouncing scale factor $a_2$. The function $a_2$ falls from a greater value at the beginning, bounce the minimum value, and then rise again at the end. Other scale factors $a_{1,3}$ start dynamics from zero. At the beginning $t=0$, the Universe is an infinite straight line along axis $x_2$: $a_1(0)=a_3(0)=0$, $a_2(0) =+\infty$.  The model may be applicable to the early Universe before the end of primary inflation. In this case, the Universe begins to evolve from the phase of the magnetic field dominance.

\subsection{A massless scalar field}

Here we will consider a massless scalar field, $V=0$. In this case from (\ref{VPsi}) we get
\begin{equation}\Psi=\Psi_0e^{-2\gamma\phi}\,.\end{equation}
Then
\begin{equation}\label{maslesH}H^2=\frac{8\pi\Psi_0c_0}{9}\left(2+\frac{8\pi }{\gamma^2}\right)\cdot\frac{1}{a^6}
\,\, \,\text{and} \end{equation}
\begin{equation}\label{maslesa}a=\left[8\pi\Psi_0c_0\left(2+\frac{8\pi }{\gamma^2}\right)\right]^{1/6}\cdot t^{1/3}\,,\,\,t\geq 0\,.\end{equation}

Function $\sigma^2/H^2$ takes the form
\begin{equation}\frac{\sigma^2}{H^2}=
\left(\frac{\frac{8\pi}{\gamma^2}-1}{\frac{8\pi}{\gamma^2}+2}\right)^2+\frac{3}{\left(2+\frac{8\pi }{\gamma^2}\right)^2}\cdot
\ln\left(\left[8\pi\Psi_0c_0\left(2+\frac{8\pi }{\gamma^2}\right)\right]^{1/2}\cdot t\right)\,.\end{equation}
 There is a time interval $t<t'$ for which $\sigma^2/H^2<0$. The anisotropy increases indefinitely as the Universe expands. From this position, this model is not interesting to us.

\section{Models with controlled anisotropy}

Due to the nonlinearity of the equations (\ref{aniz1}) and (\ref{aniz2}), the process anisotropization is possible \cite{SushkovStar, Tahara}.
How to avoid unlimited growth of anisotropy at late times? What should be the potential $V(\phi)$ and the function $\Psi(\phi)$?
Let's explore the different possibilities.

\subsection{A constant anirsotropy model}

A possible option is to require a constant level of anisotropy $\sigma/H$=const, using the last degree of freedom. Then from (\ref{sigmanaH}) it follows
$$\frac{1}{Ha^{3}}
\cdot\int dt\, \frac{e^{2\gamma\phi}\Psi(\phi)}{a^3}=\text{const},\,\, \text{or}$$
\begin{equation}\label{conssigH}\frac{1}{Ha^{3}}
\cdot\int da\, \frac{e^{2\gamma\phi}\Psi(\phi)}{Ha^4}=\text{const}\,.\end{equation}

System (\ref{phs1}), (\ref{VPsi}), (\ref{dotphi}), (\ref{conssigH}) gives functions $V(\phi)$ and $\Psi(\phi)$:
\begin{equation}\label{consV}V=Ae^{B\phi},\,\, A>0,\,\, B\neq-2\gamma\left(2+\frac{8\pi }{\gamma^2}\right),\end{equation}
\begin{equation}\label{consPsi}\Psi=-\frac{2ABc_1^6 }{c_0\left[B+2\gamma\left(2+\frac{8\pi }{\gamma^2}\right)\right]}
\exp\left\{\phi\left[B+2\gamma\left(1+\frac{8\pi }{\gamma^2}\right)\right]\right\}\,,\end{equation}
for which the anisotropy is constant.
The Hubble parameter and the scale factor are, respectively, given by
\begin{equation}H^2=\frac{32\pi A\left(2+\frac{8\pi }{\gamma^2}\right)^2}{9\left(B/\gamma+2\left(2+\frac{8\pi }{\gamma^2}\right)\right)}\left(\frac{c_1}{a}\right)
^{-\frac{3 B}{\gamma}\cdot\frac{1}{2+\frac{8\pi}{\gamma^2 }}}\,,\end{equation}
\begin{equation}\label{consscalar} a=c_1\left[-\frac{B}{\gamma} \,\sqrt{\frac{8\pi A}{B/\gamma+2\left(2+\frac{8\pi }{\gamma^2}\right)}}\,\,\cdot t\right]^{-\frac{2\gamma}{3 B}\left(2+\frac{8\pi}{\gamma^2}\right)}\,.\end{equation}

Equality (\ref{1Psi})  gives the condition
\begin{equation}\frac{B/\gamma}{B/\gamma+2\left(2+\frac{8\pi }{\gamma^2}\right)}<0\end{equation}
and $H^2>0$:
\begin{equation}\frac{1}{B/\gamma+2\left(2+\frac{8\pi }{\gamma^2}\right)}>0.\end{equation}
From the last two inequalities it follows
\begin{equation}\label{consnerav}-2\left(2+\frac{8\pi }{\gamma^2}\right)<B/\gamma<0.\end{equation}

Magnitude $\sigma^2/H^2$ takes the form
\begin{equation}\frac{\sigma^2}{H^2}=
\left(\frac{\frac{8\pi}{\gamma^2}-1}{\frac{8\pi}{\gamma^2}+2}\right)^2-
\frac{3B/\gamma}{\left(2+\frac{8\pi }{\gamma^2}\right)^2\left[B/\gamma+2\left(2+\frac{8\pi }{\gamma^2}\right)\right]}\,.\end{equation}
Anisotropy is small, $|\sigma/H|\ll 1$,  if
\begin{equation}\label{consnerav1}\left|\frac{B}{\gamma}\right|\ll 1,\end{equation}
which is consistent with (\ref{consnerav}).

Constraints (\ref{consnerav}) and (\ref{consnerav1}) determine the properties of scale factor (\ref{consscalar}):
\begin{equation}\label{consscalar1} a=c_1\left[\left|\frac{B}{\gamma}\right| \,\left\{\frac{8\pi A}{\left|B/\gamma+2\left(2+\frac{8\pi }{\gamma^2}\right)\right|}\right\}^{1/2}\,\,\cdot t\right]^{\left|\frac{2\gamma}{3 B}\left(2+\frac{8\pi}{\gamma^2}\right)\right|}\,,\end{equation}
where
\begin{equation}\label{consscalar12} t\geq 0\,\,  \text{and}\,\, \left|\frac{2\gamma}{3 B}\left(2+\frac{8\pi}{\gamma^2}\right)\right|\gg 1.\end{equation}
Therefore, the Universe is expanding with acceleration. In this model we have power-law inflation. This property is true in all directions:
\begin{equation}a_{1,2}(a)=\text{const}_{1,2}\cdot a^{k_{1,2}},\,\, a_3(a)=\text{const}\cdot a^{\frac{3}{2+\frac{8\pi}{\gamma^2}}}\,,\end{equation}
where
$$k_{1,2}=\frac{3}{2}\cdot\frac{1+\frac{8\pi}{\gamma^2}}{2+\frac{8\pi}{\gamma^2}}\pm$$
\begin{equation}\label{h12c}\pm\sqrt{
\frac{9}{4}\left(\frac{\frac{8\pi}{\gamma^2}-1}{\frac{8\pi}{\gamma^2}+2}\right)^2+\frac{9|B/\gamma|}{\left(2+\frac{8\pi }{\gamma^2}\right)^2\left[B/\gamma+2\left(2+\frac{8\pi }{\gamma^2}\right)\right]}
}\approx 1,\,\, \frac{3}{2+\frac{8\pi}{\gamma^2}}\approx 1.\end{equation}
The Universe begins its evolution from the point $a_{i}(0)=0.$  In this model, the constant parameters $\sigma/H$, $B/\gamma$ are related so that the requirement of small anisotropy leads to the inflationary scale factor $a$ (see (\ref{consnerav1}), (\ref{consscalar1}) and (\ref{consscalar12})). Therefore, the model is applicable in the era of primary inflation or late acceleration.

\subsection{Model with dynamic anisotropy}
Here we will assume
\begin{equation}\label{dinaniz}\frac{8\pi}{\left(2+\frac{8\pi}{\gamma^2}\right)Ha^{3}}
\cdot\int da\, \frac{c_0e^{2\gamma\phi}\Psi(\phi)}{Ha^4}=A^2_0\left(\frac{c_1}{a}\right)^n,\,\, n>0\,,\end{equation}
then during the Universe expansion  the anisotropy will decrease and tend to a small value:
\begin{equation}\label{dinsigmnaH}\frac{\sigma^2}{H^2}=
\left(\frac{\frac{8\pi}{\gamma^2}-1}{\frac{8\pi}{\gamma^2}+2}\right)^2+A^2_0\left(\frac{c_1}{a}\right)^n,\,\,\frac{\sigma^2}{H^2}\rightarrow \left(\frac{\frac{8\pi}{\gamma^2}-1}{\frac{8\pi}{\gamma^2}+2}\right)^2\ll1 ,\,\, \text{as}\,\, \, a\rightarrow +\infty.\end{equation}

System (\ref{phs1}), (\ref{VPsi}), (\ref{dotphi}), (\ref{dinaniz}) gives functions $V(\phi)$ and $\Psi(\phi)$:
$$V=\frac{\Psi_0c_0}{2c_1^6(3-n)}\left(ne^{-\gamma\phi n\left(2+\frac{8\pi }{\gamma^2}\right)/3}+\frac{9}{A^2_0\left(2+\frac{8\pi }{\gamma^2}\right)^2}\right)\times$$\begin{equation}\label{dinV}\times\left[e^{-\gamma\phi n\left(2+\frac{8\pi }{\gamma^2}\right)/3}+\frac{3}{A^2_0\left(2+\frac{8\pi }{\gamma^2}\right)^2}\right]^{3(2-n)/n}\,,\,\, 0<n<3\,,\end{equation}
\begin{equation}\label{dinPsi}\Psi=\Psi_0e^{\gamma\phi\left[2\left(1+\frac{8\pi }{\gamma^2}\right)-n\left(2+\frac{8\pi }{\gamma^2}\right)/3\right]}\left[e^{-\gamma\phi n\left(2+\frac{8\pi }{\gamma^2}\right)/3}+\frac{3}{A^2_0\left(2+\frac{8\pi }{\gamma^2}\right)^2}\right]^{3(2-n)/n}\,.\end{equation}

The Hubble parameters have the form:
\begin{equation}\label{dinh}H^2=\frac{8\pi}{3}\left(2+\frac{8\pi }{\gamma^2}\right)\frac{\Psi_0c_0}{c_1^6(3-n)}\left[\left(\frac{c_1}{a}\right)^n+\frac{3}{A^2_0\left(2+\frac{8\pi }{\gamma^2}\right)^2}\right]^{2(3-n)/n}\,,\end{equation}
\begin{equation}\label{dinh12} H_{1,2}=H\left[\frac{3}{2}\cdot\frac{1+\frac{8\pi}{\gamma^2}}{2+\frac{8\pi}{\gamma^2}}\pm\sqrt{\frac{9}{4}\cdot
\left(\frac{\frac{8\pi}{\gamma^2}-1}{\frac{8\pi}{\gamma^2}+2}\right)^2+3A^2_0\left(\frac{c_1}{a}\right)^n}\,\,\right]\,.\end{equation}
Parameter $H_3$ remains the same (\ref{ha1}).

Scale factors $a_{i}$ are expressed through the scale factor $a$:
$$a_{1,2}=s_{1,2}\cdot \left(\frac{a}{c_1}\right)^{\frac{3}{2}\cdot\frac{1+\frac{8\pi}{\gamma^2}}{2+\frac{8\pi}{\gamma^2}}}\left[\sqrt{1+\frac{\nu^2 }{3A^2_0}\left(\frac{a}{c_1}\right)^n}+\sqrt{\frac{\nu^2}{3A^2_0}\left(\frac{a}{c_1}\right)^n}\,\right]^{\pm\frac{2|\nu|}{n}}\times$$
\begin{equation}\label{dina12}\times\exp\left\{\mp\frac{2}{n}\cdot\sqrt{\nu^2+3A^2_0\left(\frac{c_1}{a}\right)^n}\,\right\},\,\, \nu^2\equiv\dfrac{9}{4}\cdot
\left(\dfrac{\frac{8\pi}{\gamma^2}-1}{\frac{8\pi}{\gamma^2}+2}\right)^2\,,\end{equation}
\begin{equation}a_3=s_3\cdot \left(\frac{a}{c_1}\right)^{\frac{3}{2+\frac{8\pi}{\gamma^2}}},
\,\, s_{i}=\text{const},
\,\, s_1s_2s_3=c_1^3\,.\end{equation}

Let's consider the "geometric mean"\, behavior of the Universe.
For $a/c_1\ll 1$ we have the approximation
\begin{equation}H^2\propto a^{-2(3-n)}\Rightarrow a\sim t^{\frac{1}{3-n}},\,\, t\geq 0.\end{equation}
Since $0<n<3$, then the Universe is expanding. In case $2<n<3$, the Universe is expanding with acceleration.

For $A_0^{-2/n}a/c_1\gg 1$ we have the approximation $H^2\approx$ const, which corresponds to de Sitter type expansion, $a=\text{const}\cdot e^{H_0 t}$.
In case  $0<n<2$, the phase of the Universe expansion without acceleration is replaced by a phase of accelerated expansion. When $2<n<3$, the power-law inflation is replaced by the exponential inflation over time.

\begin{figure}[h]
\includegraphics[width=10cm]{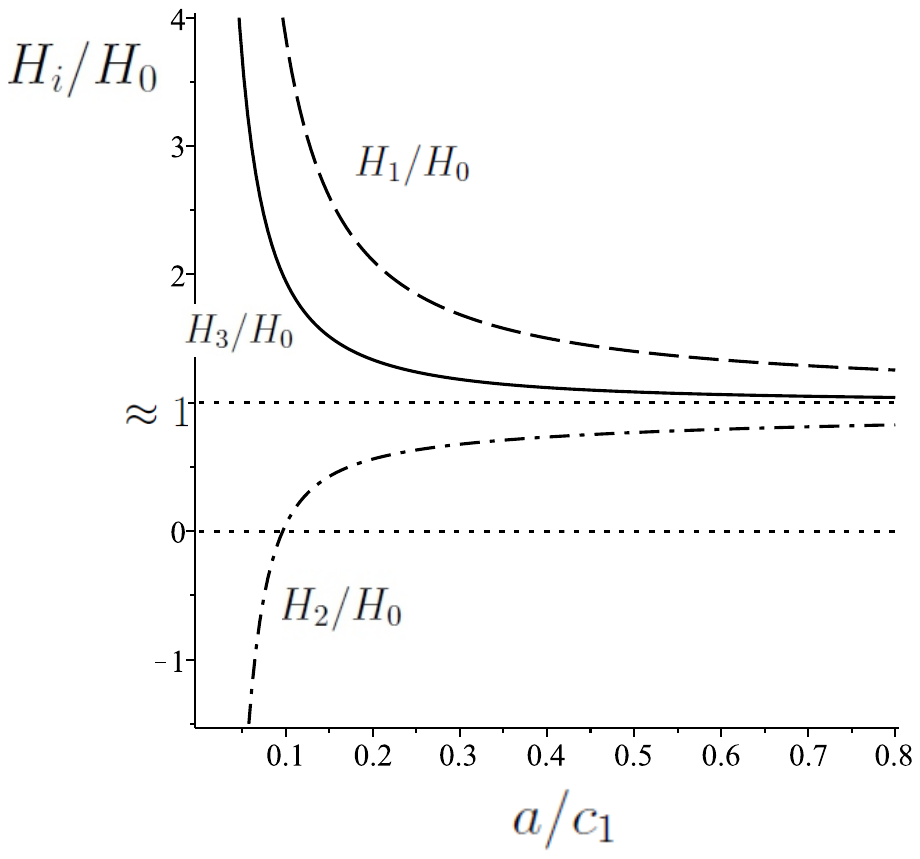}
\caption{The profile of the $H_i/H_0$; $n=3/2$, $A^2_0=10^{-2}$, $\mu=10^{-4}$. \label{risdinH123}}
\end{figure}

For example, let's put $n=3/2$, then
\begin{equation}\label{dinhn3}H^2=\frac{16\pi}{9}\left(2+\frac{8\pi }{\gamma^2}\right)\frac{\Psi_0c_0}{c_1^6}\left[\left(\frac{c_1}{a}\right)^{3/2}+\frac{3}{A^2_0\left(2+\frac{8\pi }{\gamma^2}\right)^2}\right]^{2}\,.\end{equation}
Therefore,
\begin{equation}\label{factordin}a=c_1\left\{3^{-1}A^2_0\left(2+\frac{8\pi }{\gamma^2}\right)^2\left[\exp\left(\frac{3}{2}H_\infty t\right)-1\right]\right\}^{2/3}\,,\end{equation}
$$H_\infty\equiv\frac{4\sqrt{\pi\Psi_0c_0}}{c_1^3A^2_0\left(2+\frac{8\pi }{\gamma^2}\right)^{3/2}},\,\, t\geq 0 \,.$$
We have the following approximations:
\begin{equation}\label{rem2} a\propto t^{2/3} ,\,\, \text{as}\,\, \, H_\infty t\ll 1; \,\,\, a\propto e^{H_\infty t} ,\,\, \text{as}\,\, \, H_\infty t\gg 1. \end{equation}
The first approximation corresponds to the dark matter with the equation of state parameter $w=0$. The second approximation corresponds to the dark energy with $w=-1$. Thus, a model of a unified description of the dark sector is obtained.

\begin{figure}[h]
\includegraphics[width=10cm]{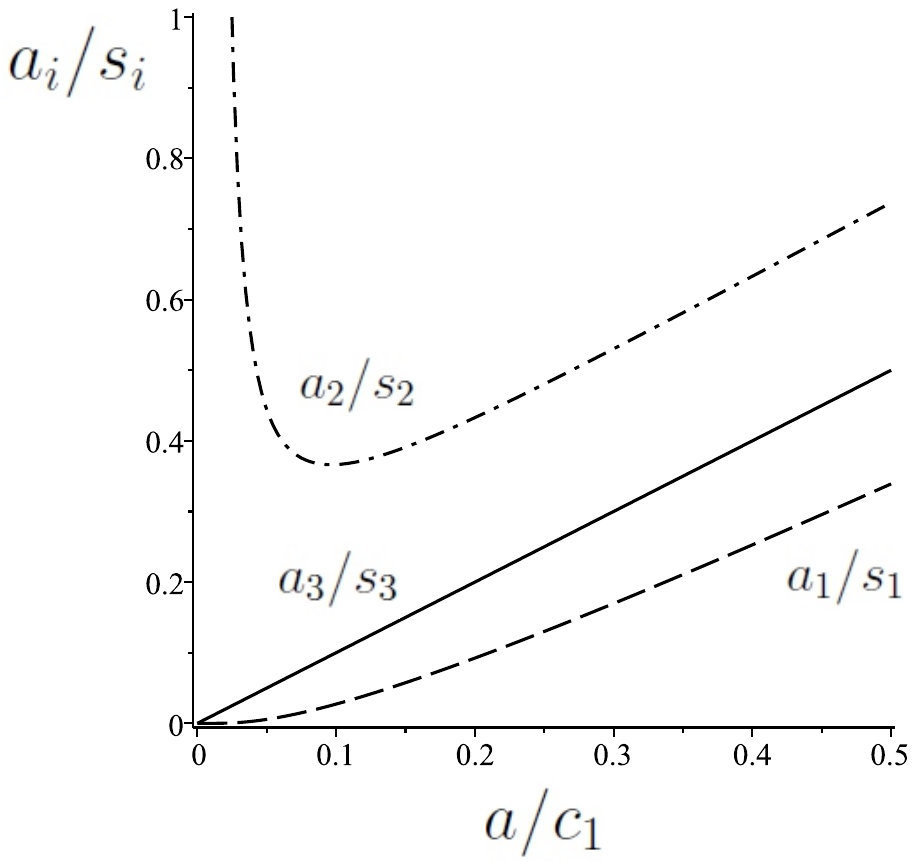}
\caption{The profile of the $a_i/s_i$; $n=3/2$, $A^2_0=10^{-2}$, $\mu=10^{-4}$. \label{risdinA123}}
\end{figure}

Next, we will consider the anisotropic properties of the model. As seen in Fig. \ref{risdinH123}, the functions $H_i$ tend monotonically to  constant values over time, $A_0^{-2/n}a/c_1\gg1$:
\begin{equation}\label{rem2} H_1\rightarrow3H_0\cdot\frac{8\pi}{\gamma^2\left(2+\frac{8\pi}{\gamma^2}\right)};\,\,
H_2, H_3\rightarrow3H_0\cdot\frac{1}{2+\frac{8\pi}{\gamma^2}}\,,\end{equation}
$$H_0=\left[\frac{8\pi}{3}\left(2+\frac{8\pi }{\gamma^2}\right)\frac{\Psi_0c_0}{c_1^6(3-n)} \right]^{1/2}\left[\frac{3}{A^2_0\left(2+\frac{8\pi }{\gamma^2}\right)^2}\right]^{(3-n)/n}\,,$$
that is, in this approximation we obtain model (\ref{bezMh123}).  According to (\ref{rem2}) and (\ref{dinsigmnaH}), the decrease in anisotropy occurs quickly, according to the exponential law.

The Universe is expanding along the $x_1$ and $x_3$ axes: $H_{1,3}>0$.
The Hubble parameter $H_2$ takes on negative and positive values. The Universe contracts in early times, and expands in late times along axis $x_2$.
For $A_0^{-2/n}a/c_1\ll 1$  scalar factors $a_{1,2}$ have the approximation:
\begin{equation}a_{1,2}(a)\sim \left(\frac{a}{c_1}\right)^{\frac{3}{2}\cdot\frac{1+\frac{8\pi}{\gamma^2}}{2+\frac{8\pi}{\gamma^2}}}
\exp\left\{\mp\frac{2\sqrt{3}A_0}{n}\cdot\left(\frac{c_1}{a}\right)^{n/2}\,\right\}.\end{equation}

Scale factors $a_i(a)$ are shown in Fig. \ref{risdinA123}. The figure shows a bouncing scale factor $a_2$. The function $a_2$ falls from a greater value at the beginning, bounce the minimum value, and then rise again at the end. Other scale factors $a_{1,3}$ start dynamics from zero. At the beginning $t=0$, the Universe is an infinite straight line along axis $x_2$: $a_1(0)=a_3(0)=0$, $a_2(0) =+\infty$.  The model may be applicable to the early Universe before the end of primary inflation.

\section{Conclusion}

We have proposed  the designer method in the BI for the subclass of the HG:
\begin{equation}G_4=1/(16\pi),\,\, G_5=-\frac{1}{32\pi \gamma X}\end{equation}
with arbitrary functions $G_2(X,\phi)$, $G_3(X,\phi)$ and with the non-minimal interaction  by the law $\Psi(\phi)F_{\mu\nu}F^{\mu\nu}$.
Here this method was applied in special case
\begin{equation} \label{concl1} G_2=\varepsilon X-V(\phi),\,\, G_3=0. \end{equation}
Then the necessary condition for small anisotropy is
\begin{equation}\varepsilon=1, \,\,\frac{8\pi}{\gamma^2}=1+\mu,\,\, |\mu|=\text{const}\ll 1.\end{equation}
This condition is not sufficient.

In the {\bf designer} framework for theories (\ref{concl1}), there is the direct connection (\ref{VPsi}) between $V(\phi)$ and $\Psi(\phi)$.
Using it as first step, we studied cases:
\begin{enumerate}
\item Without the magnetic field: $\Psi=0\Rightarrow V=$const. The model has a constant level of anisotropy, $|\sigma/H|=$const$\ll1$. The Universe has de Sitter type expansion.
\item Minimal coupling with the magnetic field: $\Psi=$const $\Rightarrow\, V=c_1+c_2\cdot e^{c_3\cdot \phi}$, $c_i$=const. The anisotropy is limited and decreases to a small value as the Universe expands. In this model, there are two phases of the evolution of the Universe.
In the first phase, there is no acceleration, and the second one is characterized by the acceleration expansion of the Universe. There is a anisotropic bounce, $a_2$ is a bouncing scale factor.  At the beginning $t=0$, the Universe is an infinite straight line along axis $x_2$: $a_1(0)=a_3(0)=0$, $a_2(0) =+\infty$.
\item The massless scalar field: $V=0\Rightarrow\,\Psi=\Psi_0e^{-2\gamma\phi}$. The anisotropy increases indefinitely as the Universe expands, $\sigma^2/H^2\propto \ln (t/t_*)$, $t/t_*\rightarrow+\infty$.
\end{enumerate}

We obtained the following models by limiting anisotropy to the first step:
\begin{enumerate}
  \item  The constant level of anisotropy, $1\gg|\sigma/H|$=const $\Rightarrow V=Ae^{B\phi}$ and
  $$\Psi=-\frac{2ABc_1^6 }{c_0\left[B+2\gamma\left(2+\frac{8\pi }{\gamma^2}\right)\right]}
\exp\left\{\phi\left[B+2\gamma\left(1+\frac{8\pi }{\gamma^2}\right)\right]\right\}.$$
In this model we have power-law inflation, $a(t)\sim t^{|\alpha|}$, $|\alpha|\gg1$.
  \item Decreasing anisotropy,  $\sigma^2/H^2=c_1+c_2/a^n$. Then $V$, $\Psi$ have the form (\ref{dinV}), (\ref{dinPsi}). In case  $0<n<2$, the phase of the Universe expansion without acceleration is replaced by a phase of accelerated expansion. When $2<n<3$, the power-law inflation is replaced by the exponential inflation over time. There is a anisotropic bounce, $a_2$ is a bouncing scale factor.  At the beginning $t=0$, the Universe is an infinite straight line along axis $x_2$: $a_1(0)=a_3(0)=0$, $a_2(0) =+\infty$.
\end{enumerate}

Thus, we solve the question  of regulating the anisotropic level. In a more general context, we have found a method of obtaining exact solutions for large subclass of the HG with the electromagnetic field.  In the next work, models with $U(a)\neq 0$ (see (\ref{1int})) are studied.

\end{document}